\documentclass[12pt]{iopart}

\usepackage{graphics}

\begin{document}
\title{Collective treatment of the isovector pair correlations. Boson representation.}
\author{G. Nikoghosyan}
\address{Yerevan State University, Yerevan, Armenia}

\author{E. A. Kolganova, \ D. A. Sazonov, \ R. V. Jolos} 
\address{Joint Institute for Nuclear Research and Dubna State University, 141980 Dubna, Moscow region, Russia}
\ead{jolos@theor.jinr.ru}
\vspace{10pt}
\begin{indented}
\item[]May 2019
\end{indented}

\begin{abstract}
The theoretical approach to consideration of the Hamiltonian with pairing forces using a technique of the finite boson representation is developed. It is shown that a simultaneous description of the pairing vibrational state in $^{56}$Ni
and the pairing rotational states with $T$=0 in the neighboring $N=Z$ nuclei  is possible if
the pairing Hamiltonian takes into account only isovector monopole pairing. However, the calculated energies of the pairing rotational states of $N=Z$ nuclei removed from $^{56}$Ni by 12 and more nucleons exceed significantly the experimental values. The possible reason of this discrepancy is discussed.
\end{abstract}
%
%
%
%

\section{Introduction}
\label{intro}
Pair correlations of nucleons in atomic nuclei play a decisive role in understanding of the excitation spectra of even-A nuclei, odd-even mass differences, moments of inertia of the well deformed nuclei and many other phenomena~\cite{Bohr,Belyaev,Soloviev,Zelevinsky}. Being applied to consideration of heavy nuclei with large differences in the numbers of neutrons and protons the theoretical technique has been developed to treat pair correlations in the system of a one kind of nucleons. However, an approach to treat pairing in the N$\approx$Z nuclei should include into consideration not only neutron-neutron and proton-proton but also neutron-proton pairs. It means, that, in principle, both isovector $T=1$ and isoscalar $T=0$ pair correlations should be considered~\cite{Goodman}.

As it is well known the spin-triplet channel of the neutron-proton interaction is attractive and more strong than the identical nucleon interaction. Nevertheless, nuclei observed in nature favor isovector monopole pairing \cite{PRC61,PLB480}. It has been suggested that the explanation of the suppressed spin-triplet pairing is the presence of the strong nuclear spin-orbit field \cite{PLB430}.

To clarify further a question of competetion of the spin-singlet and spin-triplet pairing we suggest to consider the relative binding energies of the $\alpha$-particle type nuclei. Their binding can be influenced not only by the n-n and p-p pairing correlations but also by the deutron-type correlations \cite{PRC61,PLB480,PLB430}.

The existence of the $\alpha$-like quartets in nuclei is an old problem \cite{Bertsch} which was investigated with different intensities for many years. One of the questions under investigations is an existence of $\alpha$-type condensate in the ground states of nuclei \cite{Bertsch,Flowers,Eichler,Gambhir,Hasegawa,Dobes,Senkov}. The Quartet Condensate Model was proposed for studing of the isovector pairing and quarteting correlations in $N$=$Z$ nuclei \cite{Sandulescu1,Sandulescu2,Negrea1,Sandulescu3,Sandulescu4,Negrea2,Sambataro,Negrea3}. The boson approximation technique has been applied to description of the pairing and quarteting correlations \cite{Baran}.

The fact that $\alpha$-particle nuclei present an interesting region of investigation for clarification of the relative role of isovector and isoscalar pairing has been discussed in \cite{Evans} where it was shown that $\alpha$-particle transfer takes advantage of both the isovector and isoscalar terms of the residual interaction.

In the present paper we consider a possibility to clarify a relative role of the isoscalar pairing considering on the same footing two types of collective states generated by pair correlation: pairing rotational and pairing vibrational states.
Following a discussion above  we restrict our consideration by the $\alpha$-particle type nuclei.

As it follows from the collective model of pairing correlations \cite{Bes1,Dussel1,Jolos1,KEA} the energies of the states of the pairing rotational bands formed by the sequence of the $N$=$Z$ double magic nucleus plus (minus) $n$  $\alpha$-particles are determined by the corresponding inertia parameter. The expression for this parameter derived for the case when several pairing modes are realized in a nucleus, including both isoscalar and isovector pair correlations, takes the form \cite{Jolos2,Dussel2,Bes2}
\begin{eqnarray}
\label{} \fl
\Im_{pair}~\sum_{T,J}|\Delta^J_T|^2,
\end{eqnarray}
where $B$ is a mass parameter and $|\Delta^J_T|^2$ is a mean square value of the pairing correlation function for pairing interaction with different isospin and angular momentum. Thus, different pairing type modes contribute into the inertia parameter of pairing rotations. Therefore, differences in the energies of the ground states of nuclei forming pairing rotational band give us an integral information on several pairing modes.

In contrast to pairing rotations  a concrete pairing vibrational state is related mainly to the concrete pairing vibrational mode: isovector with $J$=0 or isoscalar with different $J$. For instance, the well known pairing vibrational $0^+$ states of $^{56}$Ni with excitation energy $E^*$=5.004 MeV \cite{Bohr-Dubna} is excited in
$^{58}$Ni(p,t) and $^{54}$Fe($^3$He,n) reactions. It indicates that this state is generated by isovector monopole pairing. However, a mixing of the vibrational states of different origin can not be excluded. The situation is similar to the one related to the ground state rotational band and nuclear shape vibrational states. The ground state moment of inertia contains effects of both quadrupole and octupole deformations, however, quadrupole and octupole vibrational states can be distinguished by their electromagnetic transition properties, although double octupole vibrational states  have $I^{\pi}=0^+$ as the $\beta$-vibrational states. This means that if it is possible to describe both the energies of the pairing rotational and pairing vibrational excitations based on the Hamiltonian with isovector monopole pairing only, then other pairing modes do not play an essential role. The opposite situation will indicate on importance of the other pairing modes. We mention, however, that an interpretation of the results of calculations discussed above can be complicated by a possible coupling of the pairing vibrational states generated by different pairing modes. In this case mixing of different pairing modes and a level repulsion which follows from this can complicate interpretation of the results of calculations. We don't consider such a possibility in this paper.

The aim of the present paper is to develop a theoretical approach for consideration of the Hamiltonian with the pairing forces using a technique of the finite boson representation of the bifermion operators. Then, restricting a consideration by the isovector monopole pairing forces, to construct the collective Hamiltonian describing dynamics of the isovector pairing mode and to calculate the excitation spectra of the pairing rotational and vibrational states with $T$=0 and $J^{\pi}=0^+$ in nuclei around double magic $^{56}$Ni and $^{100}$Sn.

\section{Pairing interaction and the boson representation of the bifermion operators}
We restrict our consideration, for simplicity, by the Hamiltonian with a constant pairing
\begin{eqnarray}
\label{Eq1} \fl
H=H_0+H_{int},
\end{eqnarray}
where
\begin{eqnarray}
\label{Eq2} \fl
 H_0=\sum_{j,m,\tau}(E_j-\lambda)a^+_{jm\tau}a_{jm\tau},
\end{eqnarray}
\begin{eqnarray}
\label{Eq3} \fl
H_{int}=-\sum_{J,M,T,\tau} G^J_T (A^{JM}_{T\tau})^+A_{T\tau}^{JM}.
\end{eqnarray}
The pair creation operator $(A^{JM}_{T\tau})^+$ takes the form
\begin{eqnarray}
\label{Eq4}  \fl
(A^{JM}_{T\tau})^+ =\sum_j\sqrt{j+1/2}(A_{T\tau}^{JM}(j))^+,
\end{eqnarray}
where
\begin{eqnarray}
\label{Eq5} \fl
(A^{JM}_{T\tau}(j))^+ =\frac{1}{\sqrt{2}}\sum_{m,m',t,t'}C^{JM}_{jm jm'}C^{T\tau}_{1/2 t 1/2 t'} a^+_{jmt}a^+_{jm't'}
\end{eqnarray}
Above $T$ and $\tau$ denote isospin and its projection. Of course single particle states are characterized not only by the angular momentum $j$ and its projection $m$. However, for compactness, we did't indicate above the other quantum numbers. In our case this does not create any misunderstanding.
Thus, only interaction between nucleons occupying the same single particle states is taken into account in the Hamiltonian (\ref{Eq1}). We don't consider isoscalar pairing interaction of  nucleons occupying single particle states belonging to the spin-orbit doublets because of the large difference in their single particle energies.

For consideration of the pairing interaction it is convenient to distinguish single particle levels located below and above Fermi level. The former ones are denoted as $j_-$ and the later ones as $j_+$. After introduction of the particle and hole creation and annihilation operators
\begin{eqnarray}
\fl a^+_{jmt}=
\cases{c^+_{jmt}, \quad j\in j_+,\\
(-1)^{j-m+1/2-t}c_{j -m -t}\equiv \tilde{c}_{jmt}, \quad j\in j_- .}
\label{Eq6}
\end{eqnarray}
we obtain that
\begin{eqnarray}
\label{Eq7}
\fl (A^{JM}_{T\tau})^+=\sum_{j_+}\sqrt{j_+ +1/2}(A^{JM}_{T\tau}(j_+))^+ -\sum_{j_-}\sqrt{j_- +1/2}\tilde{A}^{JM}_{T\tau}(j_-),
\end{eqnarray}
where
\begin{eqnarray}
\label{Eq7}
\fl \tilde{A}^{JM}_{T\tau}(j_-)=(-1)^{J-M+T-\tau}A^{J-M}_{T-\tau}(j_-),
\end{eqnarray}
\begin{eqnarray}
\label{Eq8a}
\fl (A^{JM}_{T\tau}(j_{\pm}))^+=\frac{1}{\sqrt{2}}\sum C^{JM}_{j_{\pm}m j_{\pm}m'}C^{T \tau}_{1/2 t 1/2 t'}c^+_{j_{\pm}mt}c^+_{j_{\pm}m't'}.
\end{eqnarray}
Finally, the Hamiltonian (\ref{Eq1}) takes the form
\begin{equation} \eqalign{
\fl H = \sum_{j_+,m,\tau}(E_{j_+}-\lambda)c^+_{j_+ m t}c_{j_+ m t}
+\sum_{j_-,m,\tau}(\lambda-E_{j_-})c^+_{j_- m t}c_{j_- m t} \cr
\fl -\sum_{JMT\tau}\left(\sum_{j_+}\sqrt{j_+ +1/2}(A^{JM}_{T\tau}(j_+))^+\right.
 \left.-\sum_{j_-}\sqrt{j_- +1/2}\tilde{A}^{JM}_{T\tau}(j_-)\right) \cr
\fl \times\left(\sum_{j_+}\sqrt{j_+' +1/2}A^{JM}_{T\tau}(j_+')-\sum_{j_-'}\sqrt{j_-' +1/2}(\tilde{A}^{JM}_{T\tau}(j_-'))^+\right) ,}
\end{equation}
where the constant term is omitted.

We use below Dyson type boson representation of the bifermion operators \cite{Jolos2}. This boson representation is finite. Thus, there is no problem which appears if infinite boson expansions are used. However, Dyson type boson representation don't keep hermiticity relations if the standard boson metric is used. In principle, this problem can be resolved as it is discussed shortly below.

In order to have in the Hamiltonian the boson terms in degrees not higher than four we use below slightly different boson images for particle and hole creation and annihilation bifermion operators
\begin{equation}
\label{Eq10a}
\fl c^+_{j_+ m t}c^+_{j_+ m' t'}\rightarrow b^+_{m t, m' t'}(j_+)
 -\sum_{m_1 m_2 t_1 t_2}b^+_{m t, m_1 t_1 }(j_+)b^+_{m' t', m_2 t_2 }(j_+)b_{m_1 t_1, m_2 t_2}(j_+),
\end{equation}
\begin{equation}
\label{Eq10b}
\fl c_{j_+ m' t'}c_{j_+ m t }\rightarrow b_{m t, m' t'}(j_+),
\end{equation}
\begin{equation}
\label{Eq10c}
\fl c^+_{j_- m t}c^+_{j_- m' t'}\rightarrow b^+_{m t, m' t'}(j_-),
\end{equation}
\begin{equation}
\label{Eq10d}
\fl c_{j_- m' t'}c_{j_- m t }\rightarrow b_{m t, m' t'}(j_-)
 -\sum_{m_1 m_2 t_1 t_2}b^+_{m_1 t_1, m_2 t_2 }(j_-)b_{m' t', m_2 t_2 }(j_-)b_{m t, m_1 t_1}(j_-).
\end{equation}
\begin{equation}
\label{Eq10e}
\fl c^+_{j_{\pm} m t}c_{j_{\pm} m t }\rightarrow 2\sum_{m_1 t_1}b^+_{m t, m_1 t_1}(j_{\pm})b_{m t, m_1 t_1}(j_{\pm}),
\end{equation}
Here boson operators $ b^+_{m t, m' t'}(j)$ and $b_{m t, m' t'}(j)$
satisfy the following commutation relations
\begin{equation}
\label{Eq11}
\fl [b_{m t, m' t'}(j), b^+_{m_1 t_1, m_2 t_2}(j)]=\delta_{m m_1}\delta_{t t_1}\delta_{m' m_2}\delta_{t' t_2}-\delta_{m m_2}\delta_{t t_2}\delta_{m' m_1}\delta_{t' t_1}. \label{eq: comrel}
\end{equation}
Comparing the equations (\ref{Eq10a}) and (\ref{Eq10b}) or (\ref{Eq10c}) and (\ref{Eq10d}) we see that in the usual boson metric in which the scalar product of the boson operators is determined by the expression
\begin{equation}
\label{Eq11a}
\fl \left(b_{\alpha},b^+_{\beta}\right)=\delta_{\alpha\beta},
\end{equation}
where $\alpha\equiv j,m,t$, the boson images of the operators in (\ref{Eq10a}) and (\ref{Eq10b}) or (\ref{Eq10c}) and (\ref{Eq10d}) are not connected by the operation of the Hermitian conjugation. In practice, it does not introduces principle complexities, since the eigenvalues of the Hamiltonian are real if other approximations are not done. A transition to the boson representation keeping hermiticity relations in the usual boson metric can be realized using some nonunitary transformation
$\hat{U}$ which keeps, however, boson commutation relations. The new boson representation looks as
\begin{equation}
\label{Eq11b}
\fl (\hat{U})^{-1}\hat{O}_D\hat{U},
\end{equation}
where the operator $\hat{U}$ is related to the metric operator $\hat{F}$
\begin{equation}
\label{Eq11c}
\fl \hat{U}=(\hat{F})^{1/2}.
\end{equation}
Here the operator $\hat{F}$ determines the new boson metric
\begin{equation}
\label{Eq11d}
\fl \langle \Phi_{\beta},\Phi_{\alpha}\rangle =\left(\Phi_{\beta},\hat{F}\Phi_{\alpha}\right)
\end{equation}
so that the hermiticity relation
\begin{equation}
\label{Eq11e}
\fl \langle \Phi_{\beta},(\hat{O})^+_D\Phi_{\alpha}\rangle =\langle \hat{O}_D\Phi_{\beta},\Phi_{\alpha}\rangle
\end{equation}
is kept. Above $\hat{O}_D$ is a Dyson type boson image of the bifermion operator.

Let us introduce boson operators with the well defined angular momentum and isospin
\begin{equation}
\label{Eq12}
\fl b^{+JM}_{T\tau}(j_{\pm})=\frac{1}{\sqrt{2}}\sum_{m,m',t,t'}C^{JM}_{j_{\pm}m j_{\pm}m'}C^{T\tau}_{1/2t 1/2t'}b^+_{mt, m't'}(j_{\pm})
\end{equation}
It follows from (\ref{Eq11}) that
\begin{equation}
\label{Eq13}
\fl \left[b^{JM}_{T\tau}(j),b^{+J'M'}_{T'\tau'}(j)\right]=\frac{1}{2}\delta_{JJ'}\delta_{MM'}\delta_{TT'}\delta_{\tau\tau'}\left(1-(-1)^{J+T}\right)
\end{equation}
In terms of these operators the Hamiltonian takes the form
\begin{equation}
\label{Eq14}
\eqalign{
\fl H = \sum_{j_+}2(E_{j_+}-\lambda)\sum_{JMT\tau}b^{+ JM}_{T \tau}(j_+)b^{JM}_{T \tau}(j_+)
+\sum_{j_-}2(\lambda-E_{j_-})\sum_{JMT\tau}b^{+ JM}_{T \tau}(j_-)b^{JM}_{T \tau}(j_-) \cr
\fl -\sum_{JMT\tau}G^{J}_{T} \left(\sum_{j_+}\sqrt{j_+ +1/2}b^{+ JM}_{T \tau}(j_+)+
\sum_{j_-}\sqrt{j_- +1/2}\tilde{b}^{JM}_{T\tau}(j_-)\right) \cr
\fl \times\left(\sum_{j_+'}\sqrt{j_+' +1/2}b^{JM}_{T\tau}(j_+')+\sum_{j_-'}\sqrt{j_-' +1/2}\tilde{b}^{+JM}_{T\tau}(j_-')\right) \cr
\fl + 2  \sum_{JMT\tau}G^{J}_{T} \left(F^{JM}_{T \tau}(+) + F^{JM}_{T \tau}(-) \right)
\left(\sum_{j_+}\sqrt{j_+ +1/2}b^{JM}_{T \tau}(j_+)+
\sum_{j_-}\sqrt{j_- +1/2}\tilde{b}^{+JM}_{T\tau}(j_-)\right)
,}
\end{equation}
where
\begin{equation}
\eqalign{
\fl
 F^{JM}_{T\tau}(+) =\sum P^{J_1J_2J_3J'}_{T_1T_2T_3T'} \sqrt{j_+ +\frac{1}{2}}
 \left\{\begin{array}{ccc}j_+&j_+&J\\j_+&j_+&J_3\\J_1&J_2&J'\end{array}\right\}
 \left\{\begin{array}{ccc}1/2&1/2&T\\1/2&1/2&T_3\\T_1&T_2&T'\end{array}\right\} \cr
\fl \times \left((b^{+J_1}_{T_1}(j_+)b^{+J_2}_{T_2}(j_+))^{J'}_{T'} \ \tilde{b}^{J_3}_{T_3}(j_+)\right)^{JM}_{T\tau},
\cr
\fl
F^{JM}_{T\tau}(-) = \sum P^{J_1J_2J_3J'}_{T_1T_2T_3T'} \sqrt{j_- + \frac{1}{2}}
\times\left\{\begin{array}{ccc}j_-&j_-&J\\j_-&j_-&J_3\\J_1&J_2&J'\end{array}\right\}
\left\{\begin{array}{ccc}1/2&1/2&T\\1/2&1/2&T_3\\T_1&T_2&T'\end{array}\right\} \cr
\fl \times
\left(b^{+J_3}_{T_3}(j_-)\ (\tilde{b}^{J_1}_{T_1}(j_-)\tilde{b}^{J_2}_{T_2}(j_-))^{J'}_{T'}\ \right)^{JM}_{T\tau} }
\label{Eq16}
\end{equation}
and
\begin{equation*}
\fl
P^{J_1J_2J_3J'}_{T_1T_2T_3T'} \equiv \sqrt{(2J_1+1)(2J_2+1)(2J_3+1)(2J'+1)(2T_1+1)(2T_2+1)(2T_3+1)(2T'+1)} .
\end{equation*}
Above we have changed a sign of the boson operators $b^{+ JM}_{T \tau}(j_-)$ ($b^{JM}_{T \tau}(j_-)$). However, it does not changes the commutation relations.

\section{Isovectors monopole pairing and the corresponding collective Hamiltonian}

In this section we restrict our considaration by the isovector monopole pairing and keep in the Hamiltonian (\ref{Eq14})
only boson operators with $T=1$ and $J=0$. The Hamiltonian contains second order $H_2$ and fourth order $H_4$ terms in boson creation and annihilation operators.  Our following task is to determined the isovector pairing collective mode and construct the corresponding collective Hamiltonian. Of course a separation of the collective part from the total Hamiltonian is an approximation. However, large strength of the two-nucleon transfer between the ground states of the even-even  means that the assumption of the leading role of the collective mode for the energies of the low-lying states and the two-nucleon transfer has some physical grounds. The use of the boson representation significantly simplify a separation of the collective mode.

In order to determine a collective isovector monopole mode we apply a standard procedure of RPA, i.e. we  diagonalize the quadratic in bosons part of the Hamiltonian. As a result we obtain an expression of this part of the Hamiltonian in terms of operators which are linear  combinations of the noncollective boson operators:
$b^{+00}_{1 \tau}(j_{\pm})$, $b^{00}_{1 \tau}(j_{\pm})$. Depending on the strength of the pair correlations there are two possibilities. The first one corresponds to the case  of  absence of the static pair correlations. In this case we can use both: phonon creation and annihilation operators or the collective coordinate and momentum operators.  In the second case, when the strength of the pair correlations is sufficiently large  only collective coordinate and momentum operators can be used to diagonalize the quadratic in $b^{+00}_{1 \tau}(j_{\pm})$, $b^{00}_{1 \tau}(j_{\pm})$ part of the Hamiltonian. Along with the operators describing the collective isovector monopole mode the diagonalization procedure generate also operators describing noncollective modes. We don't keep them in the Hamiltonian, thus separating collective Hamiltonian from the total one.

The collective coordinate $z^+_{1 \tau}$ and momentum $p^+_{1\tau}$ operators are expressed in the following way in terms of
$b^{+00}_{1 \tau}(j_{\pm})$, $b^{00}_{1 \tau}(j_{\pm})$:
\begin{equation}
\label{Eq17}
\fl z^{+}_{1 \tau} = \sum_{j_+} w_{j_+} b^{+00}_{1 \tau}(j_+) +
\sum_{j_-} w_{j_-} \tilde{b}^{00}_{1 \tau}(j_-),
\end{equation}
\begin{equation}
\label{Eq18}
\fl p^{+}_{1 \tau} = - i\left( \sum_{j_+} v_{j_+} b^{00}_{1 \tau}(j_+) -
\sum_{j_-} v_{j_-} \tilde{b}^{+00}_{1 \tau}(j_-) \right).
\end{equation}
Since we have both pair addition and pair removal modes the coordinate $z_{1 \tau}$ is complex. Coefficients $ w_{j_\pm}$ and
$v_{j_\pm}$ are determined by diagonalization procedure which put the quadratic in bosons part of the hamiltonian $H_2$ in the form:
\begin{equation}
\label{Eq19}
\fl H_2 = \frac{1}{2B} \sum_{\tau} p^{+}_{1 \tau} p_{1 \tau} - \frac{1}{2} C \sum_{\tau} z^{+}_{1 \tau} z_{1 \tau}.
\end{equation}
 The coefficients $B$ and $C$ will be found below.  The coefficient $C$ can be positive or negative. The last case corresponds to the absence of the static pair correlations. From the requirements
 \begin{equation}
\label{Eq20}
\eqalign{
\fl  [p_{1 \tau}, z_{1 \tau'}] = - i \delta_{\tau \tau'} \cr
\fl  [p_{1 \tau}, p_{1 \tau'}] = [z_{1 \tau}, z_{1 \tau'}]  = 0 }
\end{equation}
we obtain the following orthonormalization relations
\begin{equation}
\label{Eq21}
\eqalign{
\fl \sum_{j_+} w_{j_+}v_{j_+} + \sum_{j_-} w_{j_-} v_{j_-} = 1, \cr
\fl \sum_{j_+} w_{j_+}^2 - \sum_{j_-} w_{j_-}^2 = 0, \cr
\fl \sum_{j_+} v_{j_+}^2 - \sum_{j_-} v_{j_-}^2 = 0.
}
\end{equation}

Finally, we get the following results. The quantities $C/4B \equiv \gamma$ and $\lambda$
are the solutions of the following equations
\begin{equation}
\label{Eq22}
\fl \frac{1}{G} =\sum_{j_+} (j_+ + 1/2) \frac{2(E_{j_+}-\lambda)}{[2(E_{j_+}-\lambda)]^2+\gamma} +
\sum_{j_-} (j_- + 1/2)\frac{2(\lambda-E_{j_-})}{[2(\lambda-E_{j_-})]^2+\gamma},
\end{equation}
\begin{equation}
\label{Eq23}
\fl 0 = \sum_{j_+} (j_+ + 1/2) \frac{1}{[2(E_{j_+}-\lambda)]^2+\gamma} -
\sum_{j_-} (j_- + 1/2)\frac{1}{[2(\lambda-E_{j_-})]^2+\gamma}.
\end{equation}

The inertia coefficient B is not determined by diagonalization because the scale of the collective coordinate is free since the coordinate $z_{1 \tau}$ is not related to any observable. The expressions for the  coefficients $ w_{j_\pm}$ and
$v_{j_\pm}$ are given in Appendix A1.

\section{Collective Hamiltonian in terms of coordinates and conjugate momenta}

Since our consideration is rather qualitative because of the schematic form of the residual interaction the calculations below are performed with some additional approximation. Since the coefficient $S_-$, which is given by the difference of the sums over $j_+$ single particle levels and $j_-$ levels, is much smaller then $S_+$ we assume further that $S_-=0$. Then $W=0$, and
\begin{equation}
\label{Eq29}
\fl  w_{j_\pm} = \frac{1}{V} \tilde{w}_{j_\pm}, \quad  \tilde{w}_{j_\pm} = \frac{1}{G^0_1 S_+} \ \frac{\sqrt{j_\pm + 1/2}}
{[2(E_{j_\pm}-\lambda)]^2+\gamma},
\end{equation}
\begin{equation}
\label{Eq30}
\fl  v_{j_\pm} = V \tilde{v}_{j_\pm}, \quad  \tilde{v}_{j_\pm} = G^0_1  \ \frac{\sqrt{j_\pm + 1/2}\cdot 2|E_{j_\pm}-\lambda|}
{[2(E_{j_\pm}-\lambda)]^2+\gamma}.
\end{equation}
In this approximation a second multiplier in the last line of (\ref{Eq14}) is proportional to $z_{1 \tau}$
\begin{equation}
\label{Eq31}
\fl  \sum_{j_+} \sqrt{j_+ + 1/2} b^{00}_{1 \tau}(j_+) +
\sum_{j_-} \sqrt{j_- + 1/2} \tilde{b}^{+00}_{1 \tau}(j_-) = V z_{1 \tau}.
\end{equation}

By analogy with the assumption $S_-=0$ we neglect below the following coefficients, which are also given by the differences of the sums over $j_+$ and $j_-$, assuming that they are equal to zero:
\begin{equation}
\label{Eq32}
\eqalign{
\fl  \sum_{j_+} \frac{\tilde{v}_{j_+}\tilde{w}^2_{j_+}}{\sqrt{j_+ + 1/2}} -
\sum_{j_-} \frac{\tilde{v}_{j_-}\tilde{w}^2_{j_-}}{\sqrt{j_- + 1/2}} \approx 0, \cr
\fl  \sum_{j_+} \frac{\tilde{v}^2_{j_+}\tilde{w}_{j_+}}{\sqrt{j_+ + 1/2}} -
\sum_{j_-} \frac{\tilde{v}^2_{j_-}\tilde{w}_{j_-}}{\sqrt{j_- + 1/2}} \approx 0, \cr
\fl  \sum_{j_+} \frac{\tilde{w}^3_{j_+}}{\sqrt{j_+ + 1/2}} -
\sum_{j_-} \frac{\tilde{w}^3_{j_-}}{\sqrt{j_- + 1/2}} \approx 0.
}
\end{equation}

Taking the approximation $S_-=0$ and (\ref{Eq32}) we assume, in fact, that the spectra of the collective states generated by
particle addition mode are coincide with those generated by particle removal mode. In other words, comparing the results of calculations with the experimental data we should take for comparison the average spectra of double magic nucleus plus and minus of $n$ pairs of nucleons.

Finally, in addition to the quadratic in the collective coordinates and momenta part of the collective Hamiltonian we obtain the following expression for the anharmonic part of the Hamiltonian which contains fourth order in collective operators terms:
\begin{equation}
\label{Eq33}
\eqalign{
\fl H_4=-3G_1\left(\sum_{j_+} \frac{\tilde{v}^2_{j_+}\tilde{w}_{j_+}}{\sqrt{j_+ + 1/2}} +
\sum_{j_-} \frac{\tilde{v}^2_{j_-}\tilde{w}_{j_-}}{\sqrt{j_- + 1/2}},\right)\sum_{\tau}z^+_{1\tau}z_{1\tau}\cr
\fl -\frac{\sqrt{3}G_1}{6}\left(\sum_{j_+} \frac{\tilde{v}^3_{j_+}}{\sqrt{j_+ + 1/2}} +
\sum_{j_-} \frac{\tilde{v}^3_{j_-}}{\sqrt{j_- + 1/2}}\right)\sum_{\tau}((z^+_1z^+_1)_0\tilde{z}_1)_{1\tau}z_{1\tau}\cr
\fl +\frac{\sqrt{15}G_1}{3}\left(\sum_{j_+} \frac{\tilde{v}^3_{j_+}}{\sqrt{j_+ + 1/2}} +
\sum_{j_-} \frac{\tilde{v}^3_{j_-}}{\sqrt{j_- + 1/2}}\right)\sum_{\tau}((z^+_1z^+_1)_2\tilde{z}_1)_{1\tau}z_{1\tau}\cr
\fl +\frac{\sqrt{3}G_1}{6}\left(\sum_{j_+} \frac{\tilde{v}_{j_+}\tilde{w}^2_{j_+}}{\sqrt{j_+ + 1/2}} +
\sum_{j_-} \frac{\tilde{v}_{j_-}\tilde{w}^2_{j_-}}{\sqrt{j_- + 1/2}}\right)\sum_{\tau}\left(2(\imath\tilde{p}^+_1(\imath p_1z^+_1)_0)_{1\tau}
z_{1\tau}-((\imath p_1 \imath p_1)_0\tilde{z}_1)_{1\tau}z_{1\tau}\right)\cr
\fl -\frac{\sqrt{15}G_1}{3}\left(\sum_{j_+} \frac{\tilde{v}_{j_+}\tilde{w}^2_{j_+}}{\sqrt{j_+ + 1/2}} +
\sum_{j_-} \frac{\tilde{v}_{j_-}\tilde{w}^2_{j_-}}{\sqrt{j_- + 1/2}}\right)\sum_{\tau}\left(2(\imath\tilde{p}^+_1(\imath p_1z^+_1)_2)_{1\tau}z_{1\tau}-((\imath p_1 \imath p_1)_2\tilde{z}_1)_{1\tau}z_{1\tau}\right)
}
\end{equation}

In order to analyze the potential energy it is convenient to separate in $z_{1\tau}$ the variables related to isospin rotational invariance and gauge invariance \cite{Bes1}:
\begin{equation}
\label{Eq34}
\fl z^+_{1\mu}=\Delta\exp(-\imath\phi)\left(D^1_{\mu 0}(\psi_1,\psi_2,\psi_3)\cos\theta
+\frac{1}{\sqrt{2}}(D^1_{\mu 1}(\psi_1,\psi_2,\psi_3)+D^1_{\mu -1}(\psi_1,\psi_2,\psi_3))\sin\theta\right).
\end{equation}
Here $D^1_{\mu k}(\psi_1,\psi_2,\psi_3)$ is the Wigner function and $\psi_1, \psi_2$ and $\psi_3$ are Euler angles in isospace.
Angle $\phi$ is related to the particle number operator $\hat{N}$
\begin{equation}
\label{Eq35}
\fl \hat{N}\equiv (\hat{A}-A_0)=\imath\frac{\partial}{\partial\phi},
\end{equation}
where $A_0$ is the number of nucleons in the basic double magic nucleus like $^{56}$Ni or $^{100}$Sn.
The fact that there are in total  six collective variables and that a transformation depends on four angular variables, suggests that there exists the intrinsic system in which deformation is characterized by only two parameters. The value of $\Delta$ characterizes the strength of the pair correlations and $\theta$ characterizes their isospin structure, however, in the intrinsic isospace.

The expression for the conjugate momentum is given in Appendix A2 \cite{Dussel2,Jolos1}.
Substituting the expressions (\ref{Eq34}) and (\ref{Eq36}) into (\ref{Eq19}) and (\ref{Eq33}) we obtain the collective Hamiltonian in terms
of the variable $z_{1\mu}$.

The potential energy $V$ depends on two invariants with respect to rotations in isospace and gauge space $\Delta^2$ and $\Delta^4\cos^2 2\theta$:
\begin{equation}
\label{Eq37}
\eqalign{
\fl V=-\left(G^2_1S_+\frac{C}{4B}+3G_1\left(\sum_{j_+} \frac{\tilde{v}^2_{j_+}\tilde{w}_{j_+}}{\sqrt{j_+ + 1/2}} +
\sum_{j_-} \frac{\tilde{v}^2_{j_-}\tilde{w}_{j_-}}{\sqrt{j_- + 1/2}}\right)\right)\Delta^2\cr
\fl +2G_1\left(\sum_{j_+} \frac{\tilde{v}^3_{j_+}}{\sqrt{j_+ + 1/2}} +
\sum_{j_-} \frac{\tilde{v}^3_{j_-}}{\sqrt{j_- + 1/2}}\right)\frac{1}{2}\left(1-\frac{1}{2}\cos^2 2\theta\right)\Delta^4
}
\end{equation}

As it is shown in \cite{Dussel2} the collective wave function is determined in the domain $0\le\theta\le 2\pi$
once it is known in $0\le\theta\le\frac{\pi}{4}$. All calculations involving $\theta$ can therefore be restricted to this smaller interval.
In the interval $0\le\theta\le\frac{\pi}{4}$ potential has a minimum at $\theta$=0.

\section{Collective states with $T$=0}
As it is indicated in the Introduction we restrict our consideration by the states with isospin $T$=0. In this case  we can omit the isospin operators in the
expression for $p^+_{1\mu}$ constructing the collective Hamiltonian in terms of coordinates $z_{1\mu}$ and conjugate momentum $p_{1\mu}$. We assume also that $\theta$-mode is rigid enough and we can put $\theta$=0 in the expression for the potential energy and  $p_{1\mu}$. Approximately, $\theta$-motion can be described as the harmonic vibrations around $\theta$=0. These
vibrations are quite rigid in the case of the static pair correlations or in the transition region where mean square value of $\Delta$ is large enough. Thus, we put $\theta$=0 in the expressions for the inertia coefficients for $\Delta$-vibrations and pairing rotations.

As the result we obtain the following expression for that part of the collective Hamiltonian $(H_2+H_4)$ (\ref{Eq30}) and (\ref{Eq39}) which contains the collective momentum
\begin{equation}
\label{Eq38}
\fl T_{kin}=\frac{1}{4G^2_1S_+}\left(-\frac{1}{\Delta}\frac{\partial}{\partial\Delta}\Delta(1+t\Delta^2)\frac{\partial}{\partial\Delta}+\frac{(1+t\Delta^2)}{\Delta^2}\hat{N}^2\right),
\end{equation}
where
\begin{equation}
\label{Eq39} \fl
t=\frac{1}{2}G^3_1S_+\left(\sum_{j_+} \frac{\tilde{v}_{j_+}\tilde{w}^2_{j_+}}{\sqrt{j_+ + 1/2}} +
\sum_{j_-} \frac{\tilde{v}_{j_-}\tilde{w}^2_{j_-}}{\sqrt{j_- + 1/2}}\right).
\end{equation}
We see from (\ref{Eq38}) that the inertia coefficient for $\Delta$-vibrations and pairing rotations depends on $\Delta$.

The relation (\ref{Eq23}) and the approximate relations $S_-$=0 and (\ref{Eq32}) means that the sums over $j_+$ single particle states and those taken over $j_-$ states are approximately equal to each other. Below we consider the sets of nuclei around the double magic ones or with closed subshells whose pairing collective states can be considered based on the approach developed above. In all cases considered below the set of $j_+$ or $j_-$ single particle states consists in one single particle level. Thus, the sums in (\ref{Eq27}), (\ref{Eq37}) and (\ref{Eq39}) are reduced to one term.

It is convenient to introduce the new variable $x$ instead of $\Delta$ which
are connected by the relation
\begin{equation}
\label{Eq42} \fl
\Delta=\frac{1}{\sqrt{t}}\sinh(x)
\end{equation}
and to exclude from the Schr\"odinger equation the first derivative.
As a result, the collective Hamiltonian with the kinetic energy term (\ref{Eq38}) and the potential energy (\ref{Eq37}) is presented as
\begin{eqnarray}
\label{Eq43}
\fl H=\frac{1}{8}G_1\left(-\frac{\partial^2}{\partial x^2}+\frac{(1+\sinh^2x)N^2}{\sinh^2x}+\frac{(1-4\sinh^2x-4\sinh^4x)}{4\sinh^2x(1+\sinh^2x)}\right.\cr
\fl \left.-16\frac{(2j+1)^2}{d}\left(1+\frac{3}{2j+1}-\frac{1}{d}\right)\sinh^2x+16\frac{(2j+1)^2}{d^2}\sinh^4x\right),
\end{eqnarray}
Above $d=G_1(2j+1)/2|E_j-\lambda|$ where $j$ denote the single particle level which exhaust the set of $j_+$ or $j_-$ single particle states.

For the calculations based on the Hamiltonian (\ref{Eq43}) it is important to know in what limits of variation of the variable $x$ the corresponding Schr\"odinger equation should be considered. The values of $x$ are related to the values of $\Delta$  by the relation (\ref{Eq42}). At the same time $\Delta$ is related to the boson creation and annihilation operators by equations
(\ref{Eq17}) and (\ref{Eq34}). It means that Pauli principle which determine the physical subspace of the boson space also put a restriction on the possible values of $x$. It is established in our calculations that the wave functions of the pairing rotational and lowest vibrational states obtained by solving the Schr\"odinger equation indeed approach to zero just at the boundary of the physical values of $x$.

\section{Solution of the Schr\"odinger equation for $T$=0 states}
In order to answer the question formulated in the Introduction: is it possible to describe both the energies of the pairing rotational and pairing vibrational states basing on the Hamiltonian including only monopole isovector pairing forces, we consider nuclei around $^{56}$Ni where there are necessary experimental data.

For comparison of the results of calculations with the experimental data the experimental energies have to be reduced to quantities which can be directly compared with the model predictions. For this, we subtract from the empirical binding energies those contributions which derive from sources other than  the isovector monopole pairing correlations \cite{Bes2}. For nuclei around a basic nucleus with $A=A_0$ and $Z=Z_0$ we define the quantity
\begin{equation}
\label{Eq45}
\fl E(A,Z,i)=-\left(B_{exp}(A,Z,i)-B_{LD}(A,Z)\right)+\left(B_{exp}(A_0,Z_0,gs)-B_{LD}(A_0,Z_0)\right)
\end{equation}
where $B_{exp}(A,Z,i)$ is the experimental binding energy associated with $i$th $J^{\pi}=0^+$ state with isospin $T$=0.  The quantities without index $i$ belong to the ground state. The quantity $B_{LD}(A,Z)$ is defined by the liquid drop mass formula which looks for the states with $T$=0 as
\begin{equation}
\label{Eq46}
\fl B_{LD}(A,Z)=b_{vol}A-17A^{2/3}-0.7Z^2\left(1-\frac{0.76}{Z^{2/3}}\right)/A^{1/3}.
\end{equation}
The parameter $b_{vol}$ is fixed so that $E(A_0-4,Z_0-2,gs)=E(A_0+4,Z_0+2,gs)$ \cite{Bohr-Dubna}. The results  for $E(A,Z,gs)$ obtained according to (\ref{Eq45}) with $A_0=56$ and $Z_0$=28 for nuclei with 20$\le A\le$ 128 are presented in Fig.1.
All energies are given in MeV. In the case of nuclei around $^{100}$Sn  where the experimental data on the
binding energies are absent we have used the results of calculations presented in \cite{molnix96}.

As it is seen in Fig.1 the isovector monopole pairing vibrational states can be considered basing not only on $^{56}_{28}$Ni, but also on $^{100}_{50}$Sn where the quantity $E(A,Z,gs)$ has minima. However consideration should be restricted by the values of $N \le 8$.  There are also minima at $^{28}_{14}$Si and $^{80}_{40}$Zr where filling of the subshells $d_{5/2}$ and $p_{1/2}$ creates a situation similar to that corresponding to the shell closing. However, these minima are not so deep as those at $^{56}$Ni and $^{100}$Sn, especially the minimum at $^{80}$Zr.

%
\begin{figure}
\resizebox{0.75\textwidth}{!}{%
  \includegraphics{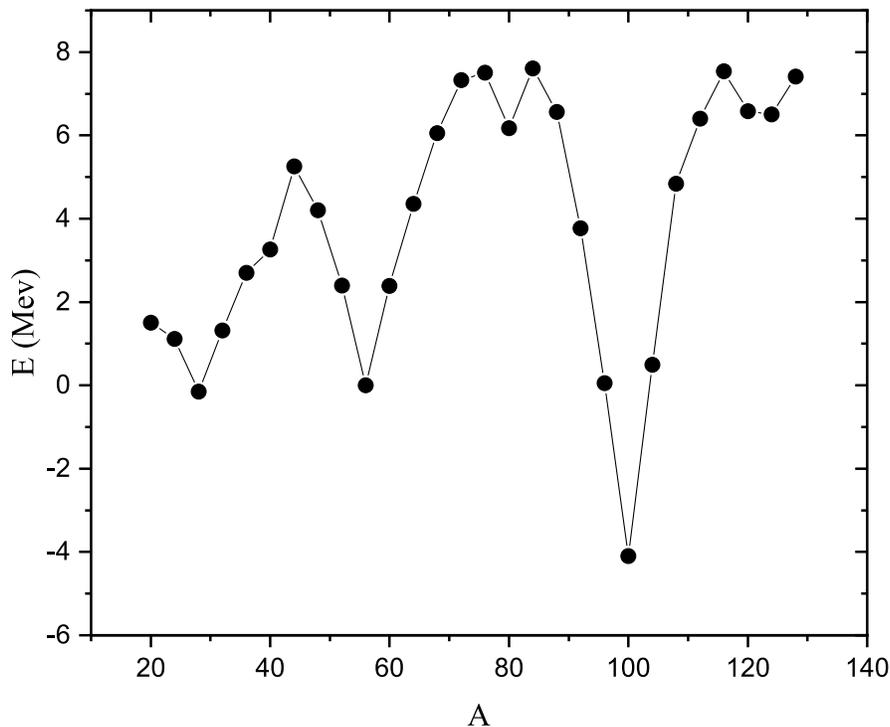}
}
\caption{The results of calculations of the energies $E(A,Z,gs)$ determined according to (\ref{Eq45}) with $A_0$=56 and $Z_0$=28 for nuclei with 20$\le A\le$ 128.}
\label{fig:1}       
\end{figure}
%
\begin{figure*}
\vspace*{5cm}       
\includegraphics{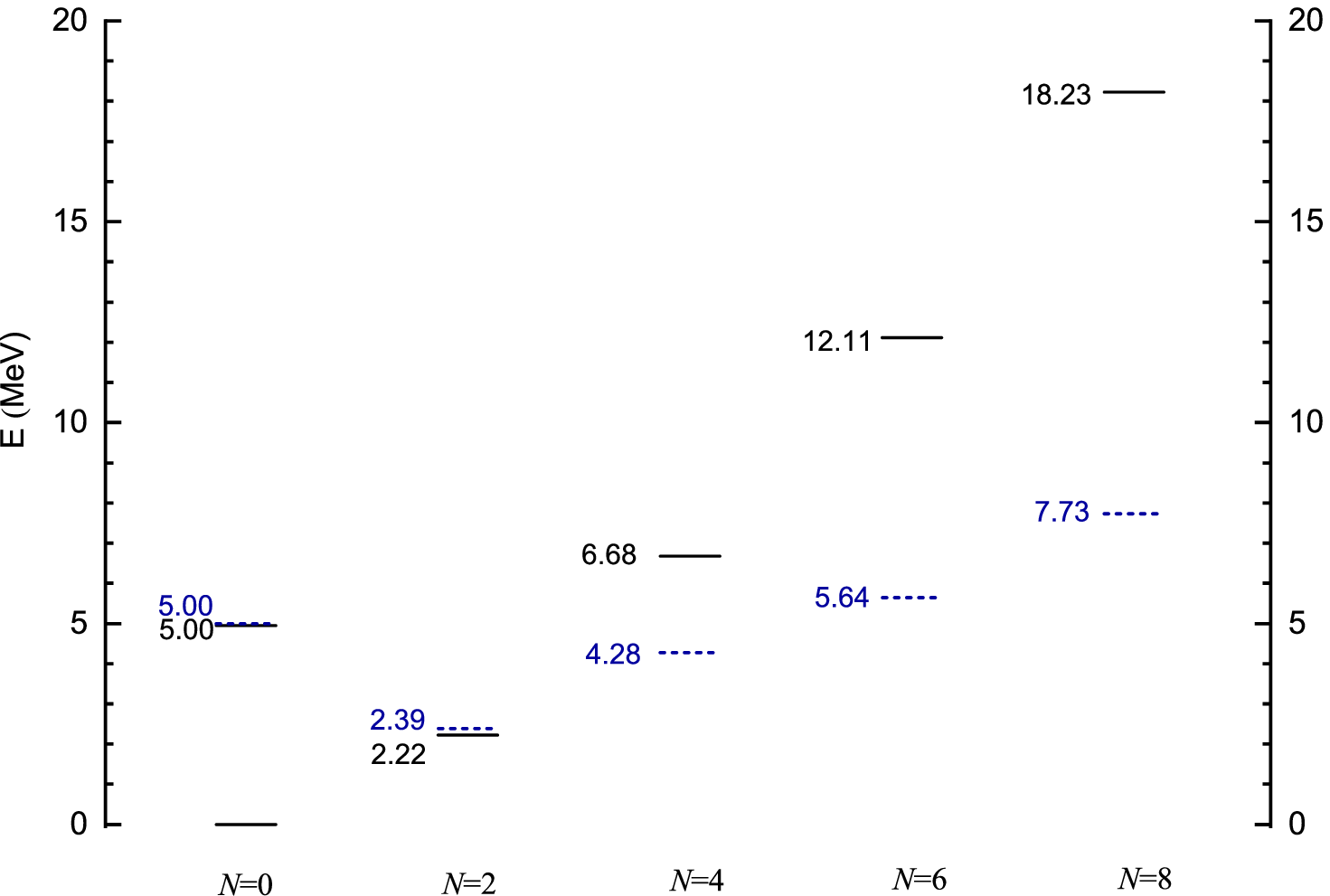}
\caption{The experimental (dashed lines) and calculated (solid lines) energies of the pairing excited states based on $^{56}$Ni. Energies are given in MeV.}
\label{fig:2}       
\end{figure*}
The results of calculations of the energies of the $0^+$ ($T=0$) ground and excited states of nuclei around $^{56}$Ni are presented in Fig.2. The value of $G_1$ has been fixed to reproduce the excitation energy of the pairing vibrational state of $^{56}$Ni at 5.00 MeV. Assuming commonly accepted nucleon number dependence of $G_1$ \cite{Soloviev1} this value corresponds to $G_1=14.4$/$A$ MeV. The parameter $d$, which characterize the ratio of the pairing strength constant to the single particle shell gap at the number of nucleons equal to 28, is determined using the value of the shell gap given by the Woods-Saxon potential. This gives $d$=0.6. The energy of the first excited state of the pairing rotational band based on $^{56}$Ni is just the energy of $^{60}$Zn ground state calculated according to (\ref{Eq45}). Our calculations give for this quantity the value 2.22 MeV. This value is close to the experimental value 2.39 MeV. The calculations have been performed also for the other values of  $d$ which has been varied near the critical value ($d_{crit}$=0.73) at which the minimum of the collective potential shifts from zero  to the nonzero value of $\Delta$. A degree of agreement with the experimental data remains, practically, the same.

Similar calculations performed for nuclei around $^{28}$Si, where the excited pairing vibrational state is known in $^{32}_{16}$S at excitation energy 5.14 MeV, also reproduce the ratio of the ground state energy of $^{32}$S, which is the energy of the first excited state of the pairing rotational band based on $^{28}$Si, to the energy of the pairing vibrational state of $^{32}$S. In this case  the values of $G_1$ and $d$ have been determined basing on the values of these parameters in $^{56}$Ni and using the $A$-dependence of $G_1$ and the single particle energy gap at $Z=A/2$=14. Thus, our calculations don't indicate on the presence of the isoscalar pairing.

At the same time we see in Fig.2 that the calculated energies of the states of the pairing rotational band formed by the lowest $0^+$ states with increasing  values of $N$ exceed significantly the corresponding experimental values starting from $N$=6. This situation is similar to that discussed by Bohr and Mottelson \cite{Scr} in connection with the states of high angular momentum $I$ in the long rotational bands. It was indicated that with increase of $I$ the possible occurrence of closed shells in a selected region of $I$ would lead to lowering of the yrast line and the enhancement of stability. In our case the number of added or removed $\alpha$-clusters play the role of the angular momentum in gauge space, and the neighboring minima of $E(A,Z,gs)$ are seen clearly in Fig 1.
Another source of the problem indicated above could be the following. Starting from the values of $N$=6 and 8 determined relative to $^{56}$Ni
it becomes unclear should these states be considered as belonging to the family of the pairing rotational states based on $^{56}$Ni or they should be considered as pairing rotational states based on $^{28}$Si or $^{80}$Zr.  The calculated value of the energy of the  state with $N$=8, determined relative to $^{56}$Ni, is 18.2 MeV. Let us consider this state as based on $^{80}$Zr. As it is seen in Fig. 1 the ground state of $^{80}$Zr is located at 6 MeV with respect to the ground state of $^{56}$Ni. The state under consideration can be treated as belonging to the addition branch of the pairing rotational state with $N=4$ determined with respect to $^{80}$Zr. Its calculated excitation energy  is equal to 5.2 MeV. Summing these two energies we obtain the energy 11.2 MeV instead of 18.2 MeV, i.e. by 7 MeV closer to the experimental value, which is 7.7 MeV. This situation is a complete analog of the situation in the Shell Model or the Interacting Boson Model where the number of the valence particles or bosons is related to the number of nucleons or holes depending on a degree of feeling of the shell. The similar result has been obtained for the pairing rotational states  based on $^{100}$Sn. Again starting from $N$=6 the calculated energies exceed significantly the experimental ones.

\section{Conclusion}
The theoretical approach for a treatment of the Hamiltonian with the pairing forces using a technique  of the finite boson representation of the bifermion operators is developed. Restricting a consideration by the isovector monopole pairing forces the excitation spectra of the pairing rotational and vibrational states with $T$=0 and $J^{\pi}=0^+$ in nuclei around double magic $^{56}$Ni and $^{100}$Sn are calculated. The calculations don't indicate on the necessity of introduction of the isoscalar pairing. It is also seen from the results of calculations that the calculated energies of the states of the pairing rotational bands formed by the nuclear ground states significantly exceed at large values of $N$ the experimental data. A possible reason of this can be an effect of the subshell closing in neighboring nuclei in which the quantity $E(A,Z,gs)$ has minima. In the case of $^{56}$Ni it can be an effect of the subshell closing in $^{28}$Si and $^{80}$Zr.

\bigskip

\appendix
\section*{Appendix A.1. Microscopic structure of the collective operators}
\renewcommand{\thesubsection}{\Alph{subsection}}


\setcounter{section}{1}
The expressions for the coefficients determining microscopic structure of the collective operators:
\begin{eqnarray}
\label{Eq24}
\eqalign{
\fl  w_{j_+} = G \sqrt{j_+ + 1/2} \ \frac{W 2(E_{j_+}-\lambda)+ V/2B}{[2(E_{j_+}-\lambda)]^2+\gamma} , \cr
\fl  w_{j_-} = G \sqrt{j_- + 1/2}  \ \frac{-W 2(\lambda -E_{j_-})+ V/2B}{[2(\lambda-E_{j_-})]^2+\gamma}, \cr
\fl  v_{j_+} =  G \sqrt{j_+ + 1/2}  \ \frac{-W C/2+ V 2(E_{j_+}-\lambda)}{[2(E_{j_+}-\lambda)]^2+\gamma} , \cr
\fl  v_{j_-} = G \sqrt{j_- + 1/2} \ \frac{-W C/2 + V 2 (\lambda -E_{j_-})}{[2(\lambda-E_{j_-})]^2+\gamma},
}
\end{eqnarray}
where
\begin{equation}
\label{Eq25}
\fl V = \frac{1}{G} \sqrt{\frac{B}{S^2_+ + S^2_- \gamma} \left(\sqrt{S^2_++S^2_-\gamma}+S_+\right)},
\end{equation}
\begin{equation}
\label{Eq26}
\fl W = \frac{1}{G} \sqrt{\frac{1}{C}\frac{1}{S^2_+ + S^2_- \gamma} \left(\sqrt{S^2_+ + S^2_-\gamma}-S_+\right)},
\end{equation}
\begin{equation}
\label{Eq27}
\fl S_+ = \sum_{j_+} (j_+ + 1/2) \frac{2(E_{j_+}-\lambda)}{([2(E_{j_+}-\lambda)]^2+\gamma)^2} +
\sum_{j_-} (j_- + 1/2)\frac{2(\lambda-E_{j_-})}{([2(\lambda-E_{j_-})]^2+\gamma)^2},
\end{equation}
\begin{equation}
\label{Eq28}
\fl S_- = \sum_{j_+} (j_+ + 1/2) \frac{1}{([2(E_{j_+}-\lambda)]^2+\gamma)^2} -
\sum_{j_-} (j_- + 1/2)\frac{1}{([2(\lambda-E_{j_-})]^2+\gamma)^2}.
\end{equation}

\section*{Appendix A.2. Collective momentum}

The expression for the conjugate momentum in terms of the variables related to the rotation in isospace and gauge space:
\begin{eqnarray}
\label{Eq36}
\fl p^+_{1\mu}\equiv-\imath\frac{\partial}{\partial z^*_{1\mu}}=\frac{1}{2}\exp{(\imath\phi)}\left(D^{1*}_{\mu 0}\cos\theta+\frac{1}{\sqrt{2}}
(D^{1*}_{\mu 1}+D^{1*}_{\mu -1})\sin\theta\right)\frac{\partial}{\partial \Delta}\cr
\fl +\frac{1}{2\Delta}\exp{(\imath\phi)}\left(-D^{1*}_{\mu 0}\sin\theta +\frac{1}{\sqrt{2}}
(D^{1*}_{\mu 1}+D^{1*}_{\mu -1})\cos\theta\right)\frac{\partial}{\partial \theta}\cr
\fl +\frac{1}{\Delta}\exp(\imath\phi)\frac{1}{\sqrt{2}}(D^{1*}_{\mu 1}-D^{1*}_{\mu -1})\left(-\sin\theta\cdot\hat{T}_0+\cos\theta\cdot\frac{1}{\sqrt{2}}(\hat{T}_1+\hat{T}_{-1})\right)\cr
\fl +\frac{1}{\Delta}\exp{(\imath\phi)}\frac{1}{2\cos 2\theta}\left(-\sin\theta\cdot D^{1*}_{\mu 0}+\cos\theta \frac{1}{\sqrt{2}}
(D^{1*}_{\mu 1}+D^{1*}_{\mu -1})\right)\cdot\frac{1}{\sqrt{2}}(\hat{T}_1-\hat{T}_{-1})\cr
\fl +\frac{1}{\Delta}\exp{(\imath\phi)}\frac{1}{2\cos 2\theta}\left(\cos\theta\cdot D^{1*}_{\mu 0}-\sin\theta \frac{1}{\sqrt{2}}(D^{1*}_{\mu 1}+D^{1*}_{\mu -1})\right)\imath\frac{\partial}{\partial\phi}.
\end{eqnarray}

\bigskip

%
%
%
%
%

\end{document}